# A new detector for the beam energy measurement in proton therapy: a feasibility study

A. Vignati[a,b,*,†], S. Giordanengo[b,*], F. Mas Milian[a,b,c,*], Z. Ahmadi Ganjeh[d], M. Donetti[e], F. Fausti[b,1], M. Ferrero[f], O. Hammad Ali[a,b,2], O. A. Martì Villarreal[a,b], G. Mazza[b], Z. Shakarami[a,b], V. Sola[b], A. Staiano[b], R. Cirio[a,b], R. Sacchi[a,b], V. Monaco[a,b].

a. Università degli Studi di Torino, Torino, Italy
b. INFN - National Institute for Nuclear Physics, Torino, Italy
c. Universidade Estadual de Santa Cruz, Ilheus, Brazil
d. Yazd University, Yazd, Iran
e. CNAO - Centro Nazionale di Adroterapia Oncologica, Pavia, Italy
f. Università del Piemonte Orientale, Novara, Italy
*These authors contributed equally to this work
[†] Corresponding author; anna.vignati@unito.it

## Abstract

Fast procedures for the beam quality assessment and for the monitoring of beam energy modulations during the irradiation are among the most urgent improvements in particle therapy.

Indeed, the online measurement of the particle beam energy could allow assessing the range of penetration during treatments, encouraging the development of new dose delivery techniques for moving targets.

Towards this end, the proof of concept of a new device, able to measure in a few seconds the energy of clinical proton beams (60 – 230 MeV) from the Time of Flight (ToF) of protons, is presented. The prototype consists of two Ultra Fast Silicon Detector (UFSD) pads, featuring an active thickness of 80 µm and a sensitive area of 3×3 mm$^2$, aligned along the beam direction in a telescope configuration, connected to a broadband amplifier and readout by a digitizer.

Measurements were performed at the Centro Nazionale di Adroterapia Oncologica (CNAO, Pavia, Italy), at five different clinical beam energies and four distances between the sensors (from 7 to 97 cm) for each energy.

In order to derive the beam energy from the measured average ToF, several systematic effects were considered, Monte Carlo simulations were developed to validate the method and a global fit approach was adopted to calibrate the system.

The results were benchmarked against the energy values obtained from the water equivalent depths provided by CNAO. Deviations of few hundreds of keV have been achieved for all considered proton beam energies for both 67 and 97 cm distances between the sensors and few seconds of irradiation were necessary to collect the required statistics.

These preliminary results indicate that a telescope of UFSDs could achieve in a few seconds the accuracy required for the clinical application and therefore encourage further investigations towards the improvement and the optimization of the present prototype.

---

[1] now at DE.TEC.TOR. Devices & Technologies Torino S.r.l., Torino, Italy
[2] now at FBK, Fondazione Bruno Kessler, Trento, Italy

## 1. Introduction

The beam energy is one of the main irradiation parameters in particle therapy, essential to target the tumor in a wide range of depths into the patient (up to 30 cm) with the required clinical range accuracy of 1 mm.

Range deviations have the potential to alter the dose distribution, leading to extreme local tumor under-dosage or normal structure over-dosage (1). The clinical effect is more significant with the widespread of the Pencil Beam Scanning (PBS) technology that usually treats fields with high dose gradients and is more sensitive to beam position, range and target misplacements (2). It is therefore of paramount importance to assert the beam range in the routine quality control checks, nowadays mainly performed by measuring the Integrated Depth-Dose (IDD) profiles in water phantoms with movable large area detector or with multi layers ionization chambers (3).

A great effort of the research community is nowadays focused on the management and reduction of range uncertainties through the development of instrumentation and methods to monitor the range of charged particles in the patient during or just after the treatment (4–8). At present, the online control of the beam transverse position is well managed with fast steering magnets and advanced beam delivery systems, able to online correct for the lateral beam deviation (9,10), while the online measurement of the beam energy and the longitudinal correction are still open issues (3). Indeed, the existing detectors (11) do not measure the beam energy during the treatment and the proper accuracy of the extracted beam energy is guaranteed by safe checks of the accelerator settings and Quality Assurance (QA) measurements of the beam (12–14).

In this scenario, a device for the direct and fast online measurement of the beam energy would be of great benefit and could be exploited for regular QA controls, for the energy check before the irradiation of new spills, and for the development of new beam monitors for future delivery schemes employing fast energy modulation systems. For instance, *beam tracking* is an organ motion mitigation technique already applied in conventional radiotherapy, aiming at adapting the beam direction to the variations of the organ position during irradiations. In charged particle therapy, the beam path variations due to longitudinal deformations caused by organ movements could be compensated with fast beam energy changes (15–19). However, the adoption of *beam tracking* techniques in particle therapy still represents a great challenge (20) and requires the development of fast energy modulation systems and, in parallel, of a fast and precise beam energy control.

A standard method used to measure the energy of particles of a monoenergetic beam consists in measuring the average Time of Flight (ToF) needed to travel a known distance between two sensors, thus obtaining the average velocity and kinetic energy (21). However, a detector suitable to measure the energy of clinical proton beams using ToF techniques should face several challenges, such as the required time resolution and the radiation hardness. Indeed, the clinical tolerance on the range uncertainty (typically less than 1 mm) corresponds to an accuracy of the energy measurement ranging from about 0.5 MeV for 230 MeV protons to 1 MeV for 60 MeV protons. This leads to a maximum acceptable error of the measured ToF ranging from 90 ps at 60 MeV to 4 ps at 230 MeV for 1 meter distance between the sensors, and these limits are more stringent for reduced flight distances. Moreover, the design and the technology of the sensors should guarantee their suitability for a clinical use, in terms of minimal perturbation of the beam, radiation resistance (one year of clinical proton irradiation in a single treatment line roughly corresponds to about $10^{15}$ protons/cm$^2$), and energy measurement within a subsecond time-frame to allow the online beam qualification.

This work describes a novel detector prototype for the measurement of the ToF of clinical protons, which exploits the advantages of the Ultra Fast Silicon Detector (UFSD) technology, recently established in High Energy Physics experiments (22,23).

The preliminary results obtained with the clinical beam of the Centro Nazionale di Adroterapia Oncologica (CNAO, Pavia, Italy) will be reported, demonstrating the feasibility of the proposed novel detector for the fast and reliable assessment of the beam energy in protontherapy.

## 2. Methods

### 2.1. UFSD technology

The proposed detector prototype for the accurate ToF measurement of clinical proton beams is made of a telescope of two UFSD sensors placed at a specific distance between each other and aligned along the beam direction.

UFSDs are n-in-p silicon sensors based on the Low Gain Avalanche Diode (LGAD) technology (24), featuring an internal moderate gain due to a thin p+ additional layer located below the n++ electrode of a heavily doped junction (22). The doping profile is characterized by a large increase in doping concentration in close proximity to the junction, resulting in a local increase of the electric field (Figure 1). The electrons produced by the incoming ionizing radiation and drifting towards the n++ electrode are accelerated in the high electric field region, generating additional hole-electron pairs. The charge multiplication mechanism results in an enhanced signal with a controlled gain value of ~10÷20, which allows retaining the linearity with the charge produced in the depleted region and keeping a low dead time. The Signal-to-Noise (S/N) ratio can be controlled by changing the voltage bias, allowing to optimize the separation of the signal from the noise and to compensate for gain reductions due to radiation damage.

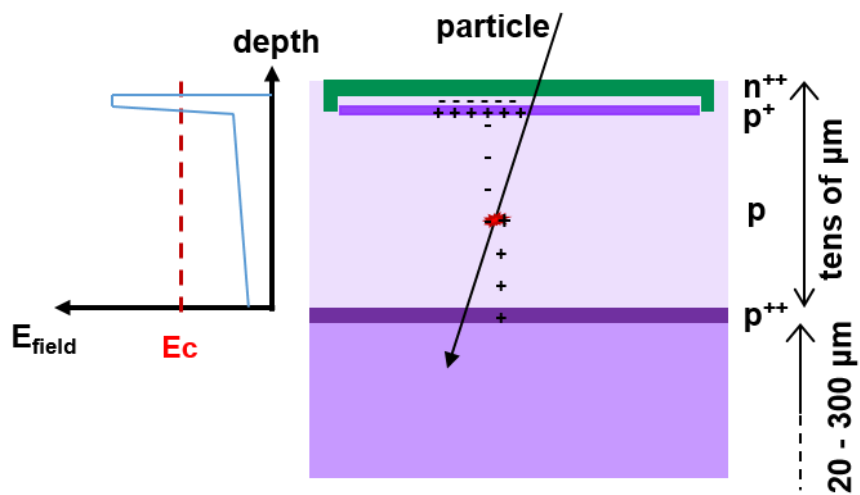

**Figure 1**. Schematic of a UFSD: the extra deep p+ layer implanted under the n++ cathode creates a strong electric field ($E_{field}$). When the electric field exceeds a critical value (Ec ~300 kV/cm) it generates charge multiplication. The electric field behaviour as a function of the sensor depth is shown on the left, while increasing tones of purple reflect the increasing p-doping concentration in the sensor. The detail of the creation and drifting of an electron-hole pair is also shown as an example. The typical thickness of the depleted region is tens of μm, whereas the underlying handling support, typically 300 μm thick, can be safely thinned down to tens of microns after sensor manufacturing.

The main advantage of a UFSD is to provide an enhanced signal with a fast rising edge in thin detectors, leading to a time resolution of about 30 ps in 50 μm active thickness (23,25). This allows concurrently accurate measurements of time and space in segmented sensors, and guarantees a reduced perturbation of the beam, as the properly thinned down thickness reduces the multiple scattering effects.

*2.2. Telescope system and readout*

A telescope prototype for the ToF measurement was built using two UFSDs (named $S_1$ and $S_2$ in the following) manufactured by Hamamatsu Photonics K.K. (HPK, Figure 2). Each sensor is segmented in four pads, each one characterized by a sensitive area of 3×3 mm$^2$ and a total thickness of silicon crossed by beam particles of 230 μm (80 μm active and 150 physical thickness).

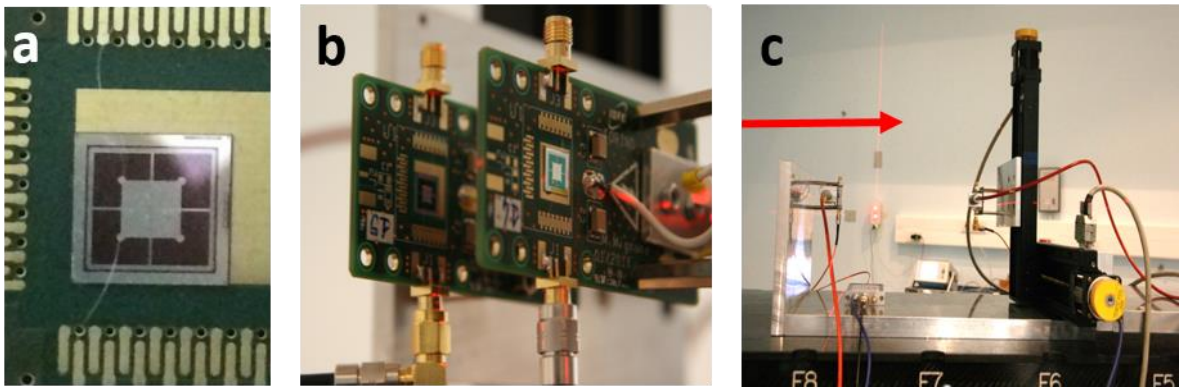

**Figure 2**. a) An HPK UFSD segmented in 4 pads; b) the telescope of two UFSDs mounted on 2 channels High Voltage (HV) distribution boards; c) the mechanical support for the detector $S_2$, equipped with two orthogonal moving stages. The red arrow in c) points out the beam direction.

The two UFSDs were mounted on general purpose High Voltage (HV) distribution boards, where only one out of the four pads was bonded to the signal output connector mounted on the board (Figure 4 a,b). The HV was set independently on the two sensors via the board connection to an external power supply (26).

Each of the two signals of the sensors were fed into a low-noise current amplifier (27) featuring an analogue bandwidth of 2 GHz and a 40 dB gain, and acquired by a 16+1 channels digitizer desktop module (28). The digitizer samples the signal at 5 GS/s, with one ADC count corresponding to 0.24 mV, and for each trigger stores 1024 samples corresponding to a waveform of 204.8 ns duration. A PC, connected to the digitizer with an 80 MB/s optical link, was used to control the acquisition, collect the waveforms, and to produce an asynchronous software trigger when the previous event was stored in memory. The conversion time of the digitizer (110 µs) and the time needed to transmit and store the data (~500 µs) limit the trigger rate to about 1.67 kHz, corresponding to an acquisition efficiency of 0.4 per-mille.

The mechanical system of the ToF telescope, shown in Figure 2c, consists of a rigid horizontal support with 10 grooves for precise positioning of $S_2$ at ten different distances from $S_1$, the latter being kept at a fixed position in the isocenter of the treatment room. In order to align the two sensors with the beam direction, two movable stages were used to support and remotely move $S_2$ in two orthogonal directions transversely to the beam.

### 2.3. Detector requirements

The accuracy of the average ToF measurement depends on the irradiation time needed to acquire the proper statistics and on the possible source of uncertainties in the identification of the coincident proton signals in the two sensors of the telescope.

For a clinical beam energy detector, the ToF precision is constrained by the required resolution of the energy measurement, which ranges from about 0.5 MeV for 230 MeV protons to 1 MeV for 60 MeV protons. These values correspond to the clinical tolerance in the range uncertainty of less than 1 mm in water, at the therapeutic proton energies (60 – 230 MeV). To meet this goal, the maximum ToF uncertainty scales with the distance between the sensors of the telescope system and decreases with the increasing particle energy. Figure 3 shows the maximum uncertainty on the ToF per unit flight distance corresponding to an uncertainty on range in water < 1 mm as a function of the proton kinetic energy, considering 500 µm uncertainty on the distance. As an example, for a distance of 1 m between the sensors, the maximum error allowed on the ToF ranges from 90 ps at 60 MeV to 4 ps at 230 MeV, and these limits are more stringent for reduced flight distances.

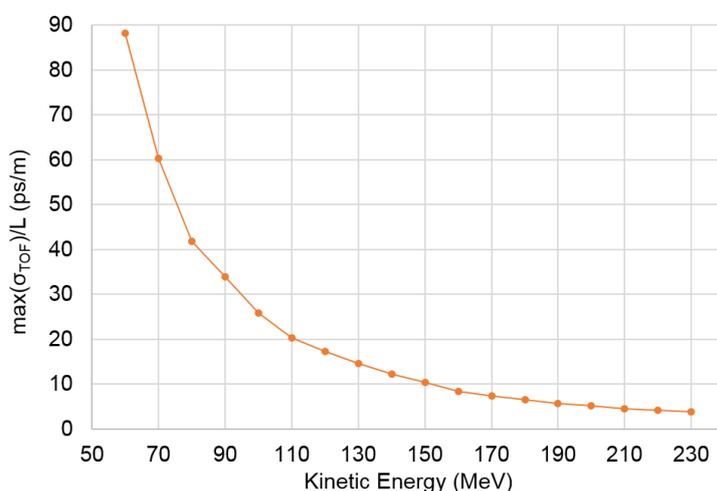

**Figure 3.** Maximum acceptable error on ToF per unit distance to measure the energy with an uncertainty corresponding to less than 1 mm error in the beam range in water. The NIST database[3] was used for the energy/range mapping.

The statistical errors depend on the duration of the beam irradiation, the consequent number of coincident proton signals acquired by the two sensors, and the relative ToF resolution. At the typical fluxes of therapeutical proton beams ($10^8 – 10^{10}$ p·s$^{-1}$·cm$^{-2}$), the data acquisition and data throughput represent the bottleneck for the reduction of the irradiation time needed to acquire the sufficient statistics.

---

[3] https://physics.nist.gov/PhysRefData/Star/Text/PSTAR.html

The ToF approach relies on the identification of coincident signals, i.e. signals generated in the two sensors of the telescope by the same proton crossing both of them. The main sources of uncertainties are the combinatorial error due to multiple protons travelling through the two sensors within the ToF frame and/or hitting only one of the two sensors due to beam divergences, to multiple scattering, and to the misalignments of the sensors with respect to the beam line. A good compromise between sensor size and travel distance must be chosen, based on simulation studies, and a careful alignment system must be employed to maximize the number of coincidences and keep the combinatorial error at acceptable levels.

The system efficiency, defined as the probability that a proton crossing the first sensor hits the second one at a specific distance along the beam trajectory, was studied with Geant4 (29) simulations for 50 µm thick sensors of different areas aligned to a beam with Gaussian transverse hape of 10 mm full-width-at-half-maximum (FWHM). This corresponds to the average pencil beam size at CNAO expressed in terms of FWHM in air at the isocenter, ranging from 7 mm at high energies up to 22 mm at lower energies (30). Figure 4 shows an example of the system efficiency for a 60 MeV proton beam at different distances between the sensors and for different sensor areas. The loss of efficiency with smaller detector area and with larger distance is due to the multiple scattering effect, which deviates the beam particles from their original trajectory after crossing the first sensor. The drop of efficiency with distance is less severe for larger energies, indicating that a minimum area of 3x3 mm$^2$ is necessary for taking measurements at the largest distance if a minimum efficiency of 10% is desired.

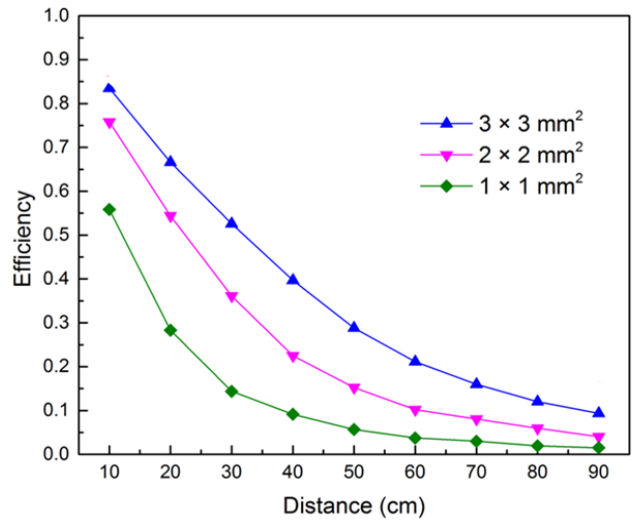

**Figure 4.** Efficiency of the telescope system for 60 MeV protons as a function of the distance between the sensors, as determined by a Geant4 simulation. A beam transverse FWHM of 1 mm and three different sensor areas (3×3, 2×2, 1×1 mm$^2$) have been considered.

### *2.4. Test with the CNAO proton beam*

To prove the detector capability of measuring the proton beam energy, the described telescope was tested in a clinical treatment room of the CNAO facility. The CNAO proton beam is provided by a synchrotron with the PBS technique (12). The ToF measurements were performed for five proton beam energies (58.9, 77.6, 103.5, 148.5, 226.1 MeV) at a beam fluence rate of $5\times10^8$ p·s$^{-1}$·cm$^{-2}$. The considered energy values (hereinafter defined *nominal energies*) were retrieved from the PSTAR dataset of the National Institute of Standards and Technology[4], corresponding to five water equivalent depths (respectively: 30, 49, 82, 155, 320 mm) measured by CNAO with a maximum deviation $< \pm 0.15$ mm (12).

Four each beam energy, the two sensors were positioned at four relative distances (7, 37, 67, 97 cm). These distances were measured with an external ruler and are not very accurate. The actual distances were obtained as a result of the calibration procedure described in Section 2.7.

For each distance $d$, a first alignment of the two sensors was done using the lasers of the treatment room (which indicate the isocenter position and the beam path), placing $S_1$ at the isocenter. Then, the $S_2$ transversal position was repeatedly changed by the movable stages covering a grid of positions with a 2 mm step, remotely set by the operator. The best position was identified as the one with the maximum number of coincidences among all grid positions, calculated as explained in the following. Once the system was aligned, the mean ToF value was measured in a dedicated synchrotron run for the selected energy values. For each run, from 5 thousands to 15 thousands waveforms were acquired from the digitizer, corresponding to a total acquisition time of around $3 - 9$ s respectively. Indeed, the beam irradiation time was increased with the increased relative distance of the sensors.

The beam energy was derived from the ToF mean value, taking into account several systematic effects and performing a system calibration, as described in the following sections, and then compared to the nominal one.

### 2.5. *Time measurement and identification of coincident signals*

An example of signals acquired during the test at CNAO, with the setup described in Sections 2.2 and 2.4, is reported in Figure 5 for a 228 MeV proton beam. The two waveforms correspond to the signals from the two sensors of the telescope, positioned at a distance of 97 cm simultaneously collected in a portion of the digitizer time window of 204.8 ns. The bias voltage of the sensors was 700 and 300 V, respectively, corresponding to an internal gain of about 5. The proton signals, each of 2 ns time duration, can be discriminated by applying a common threshold, which value is chosen by optimizing the signal separation from noise (31). The amplitude fluctuations are due to the statistical nature of the energy deposition process, while the time distribution of the particles in each detector mainly depends on the beam structure.

To minimize the dependence of the measured time on the signal amplitude, the time of arrival $t_i$ of each discriminated signal $i$ is reconstructed using the Constant Fraction Discriminator (CFD) algorithm, which associates to each proton hit the value of the time corresponding to a fixed fraction of the maximum amplitude of the signal (here set at 80% to allow the time calculation even in the presence of partially overlapping peaks).

The most probable value of the peak amplitude increases by lowering the beam energy, because of the larger energy loss in silicon. Therefore, the signals shown in Figure 5 (related to the 228 MeV beam) are examples of the smallest peak amplitudes among the signals acquired during the test at CNAO, with the fluctuations due to the statistical nature of the charge production in silicon. In this example, the coincident signals of several protons passing through both sensors can be observed, and the time difference of the coincident signals in the two sensors for each proton are the basis for the ToF measurement. Single signals without a coincidence can also be observed. When combined with the signals of the other sensor, these represent a combinatorial background, which increases with the beam flux.

---

[4] https://physics.nist.gov/PhysRefData/Star/Text/PSTAR.html

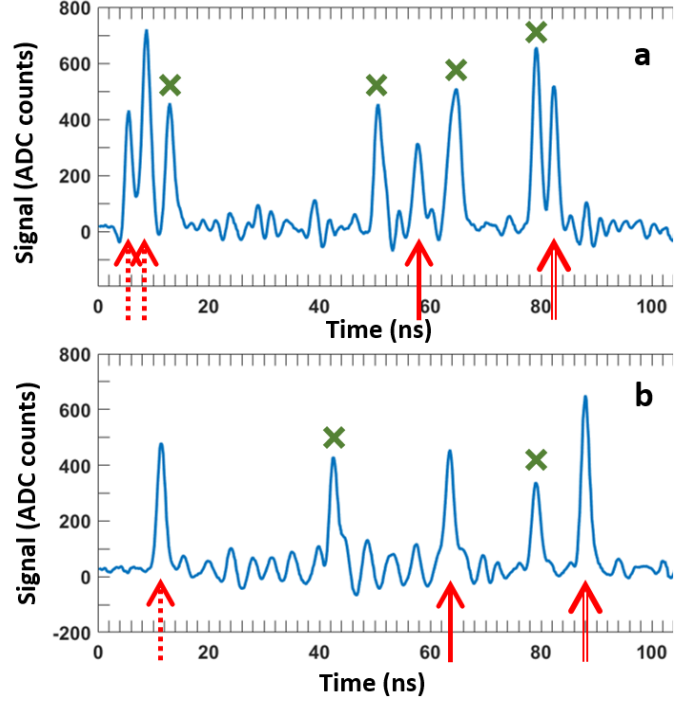

**Figure 5.** Example of two waveforms: *a)* from $S_1$ and *b)* from $S_2$ as acquired with the CNAO proton beam at 228 MeV and 97 cm distance between the two sensors of the telescope. Arrows of the same type in *a* and *b* point out the possible coincident signals in the two sensors, i.e. the two dotted arrows in *a* identify the two signals which might be in coincidence with the signal indicated by the same dotted arrow in *b*. The crosses point out the signals with no matching signal in the other sensor.

### 2.6. ToF measurement procedure

The ToF of the protons of a specific beam energy, measured with the sensors in the telescope at a specific distance, is defined as:

$$ToF = \Delta t_{mean} - offset, \qquad (1)$$

where $\Delta t_{mean}$ is the mean difference of the coincident signals times of arrival in the two sensors and *offset* is a constant time difference mainly due to the routing of the electronic chain.

The time offset is obtained with the system calibration described in Section 2.7, while the identification of the coincident signals and the measurement of the corresponding $\Delta t_{mean}$ is performed with an iterative statistical method divided in two steps, as reported in the following.

1) For a specific distance *d* between the sensors and a specific beam energy *K*, the expected mean time difference of coincident signals ($\Delta t_{est}$) is estimated from Eq. 3 reported in the following. For each peak *i* in the first sensor with time of arrival $t_{1i}$, a search of all the hits *j* in the second sensor with a time of arrival $t_{2j}$ compatible with $\Delta t_{est}$ within a search time window $\Delta t_{max}$ (10 ns) is performed ($|t_{2j}-t_{1i}-\Delta t_{est}|< \Delta t_{max}$). The peaks *i* and *j* satisfying this condition are identified as possible coincident signals and the corresponding time differences $\Delta t = t_{2j} - t_{1i}$ are reported in a histogram.

An example of the distribution of $\Delta t$ values is shown in Figure 6 for a proton beam of 103.5 MeV nominal energy and a distance of 67 cm between the sensors. The Gaussian distribution of $\Delta t$ of the true coincident signals is superimposed to an approximately flat distribution representing the combinatorial background.

2) A fit of the histogram is performed using a convolution of two Gaussian fits (red curve in Figure 6), resulting in a Gaussian curve of the peak and a second Gaussian curve of the background. The area of the first Gaussian curve is used to estimate the number of coincident signals (i.e. the number of identified protons passing through the two sensors), while $\Delta t_{est}$ is updated to its mean value ($\Delta t_{mean}$).

Both step 1, with $\Delta t_{est} = \Delta t_{mean}$ and reducing the search time window ($\Delta t_{max} = 2$ ns), and step 2 are repeated allowing to refine the $\Delta t_{mean}$ value.

An additional fit of the histogram with a single Gaussian on the central part of the distribution is then performed excluding $\Delta t$ values outside a 1.5 σ interval from the estimated $\Delta t_{mean}$ (blue curve in Fig. 6) to refine the final $\Delta t_{mean}$ value.

During the alignment phase, the just described procedure was adopted to estimate the number of coincident signals for each of the grid positions and identify the best alignment, i.e. the position with the maximum number of coincidences.

After the alignment, a dedicated acquisition is performed and the $\Delta t_{mean}$ obtained by the iterative procedure is used to determine the corresponding ToF value after subtracting the time offset obtained with the system calibration described in Section 2.7.

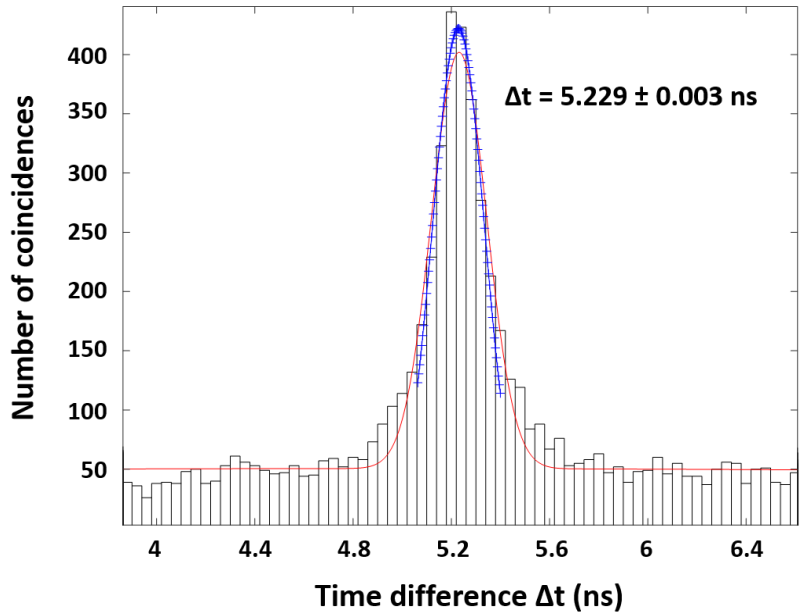

**Figure 6.** Example of $\Delta t$ distribution for a proton beam of 103.5 MeV nominal energy and a distance of 67 cm between the sensors. The red line shows the first fit (double Gaussian) and the blue line is the additional Gaussian fit for ToF values within 1.5 σ, performed to estimate the final $\Delta t_{mean}$ value.

*2.7. System calibration*

The main sources of systematic errors for the setup used during the test at CNAO reside in the uncertainty of the distance between the sensors and of the time offset related to the signal routing of the electronic chain. Assuming the nominal energies at the isocenter as known quantities, a Chi-square minimization method was adopted to calibrate the system in terms of the four positions $d_j$ between the sensors and the time offset (five free parameters, in total), using 16 ToF measurements out of the 20 performed at the CNAO test. Indeed, one proton beam energy (103.5 MeV) was not considered in the calibration method and was used to independently test the calibration results. The Chi-square was defined as

$$\chi^2(offset, d_j) = \sum_{i,j} \left\{ \frac{(\Delta t_{ij} - offset) - ToF(K_i, d_j)}{\sigma_{ToF_{ij}}} \right\}^2, \quad (2)$$

where $d_j$ (for $j=1 \div 4$) are the 4 distances and $K_i$ ($i=1 \div 4$) are the 4 energies used in the CNAO test. In Eq.(2) $\Delta t_{ij} - offset$ is the measured ToF and $ToF(K_i, d_j)$ is the expected ToF, obtained as

$$ToF(K, d) = \frac{(K + E_o) d}{c \sqrt{(K + E_o)^2 - E_0^2}}, \quad (3)$$

where $K$ is the nominal beam kinetic energy corrected for the energy loss in the first detector and in air, as explained in the following Section 2.8, and $E_0$ is the proton mass energy at rest.

Equation 2 is minimized as a function of the offset parameter and the distances $d_j$ for the 4 energy values.

*2.8. Beam energy calculation procedure*

From the ToF of protons and the known distance $d$ between $S_1$ and $S_2$, the mean kinetic energy $K$ of protons can be determined using the following equation:

$$K = E_0 \cdot \left( \frac{c \cdot ToF}{\sqrt{c^2 ToF^2 - d^2}} - 1 \right). \quad (4)$$

This formula assumes a constant velocity of the particles, thus neglecting the loss of kinetic energy in the first sensor and in the distance travelled in air between the two sensors. Under the assumption of a constant stopping power in air, it can be easily shown that the kinetic energy of Equation 4 is the kinetic energy $K_{avg}$ in the central position between the two sensors. The assumption of a constant stopping power is justified by the results of Geant4 simulations, showing that in the worst case the maximum variation of the stopping power is 1% for a 60 MeV proton along a 1 m distance.

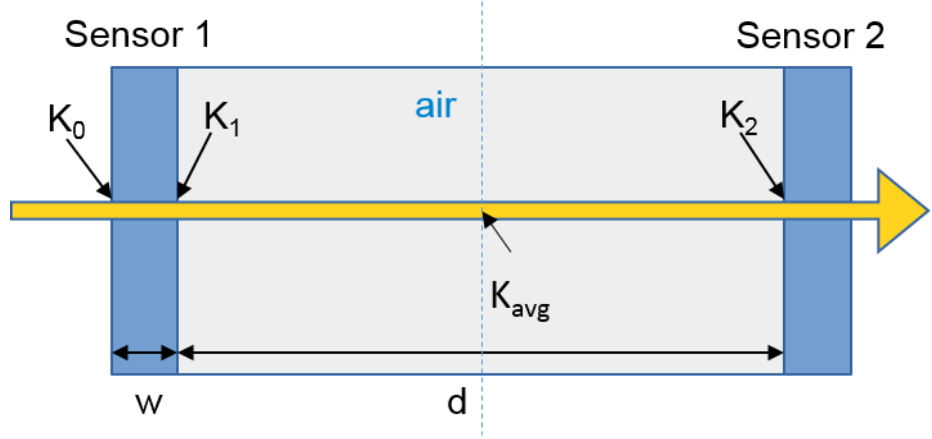

**Figure 7.** Sketch of the telescope system made by 2 UFSD sensors ($S_1$ and $S_2$ with thickness $w$) in air at a relative distance $d$. $K_0$ is the nominal proton beam kinetic energy at the entrance of the telescope. The energy loss in $S_1$ reduces to $K_1$ the initial energy $K_0$, while at the entrance of the second sensor $S_2$ the energy is reduced to $K_2$, due to the energy loss in air. The average kinetic energy $K_{avg}$ obtained from the ToF can be determined as $(K_1+K_2)/2$.

Therefore, referring to Figure 7, the proton beam energy at the isocentre $K_0$ can be obtained as

$$K_0 \approx K_{avg} + \left(\frac{S}{\rho}\right)_{air} \cdot \rho_{air} \cdot \frac{d}{2} + \left(\frac{S}{\rho}\right)_{Si} \cdot \rho_{Si} \cdot w \,, \tag{5}$$

where $\rho$ are the densities of air and silicon, respectively, $w$ is the thickness of $S_1$, $d$ is the distance between $S_1$ and $S_2$, $(S/\rho)_{air}$ is the mass stopping power in air at the energy $K_{avg}$ and $(S/\rho)_{Si}$ is the mass stopping power in silicon at the energy $K_1$, with $K_1$ obtained by summing the first two terms of Equation 5.

The PSTAR dataset has been employed for determining the value of the mass stopping power $(S_{air}/\rho_{air})$ and $(S_{Si}/\rho_{Si})$. To estimate the mass stopping power values for all the kinetic energies, the following parameterization proposed by (32) was used:

$$\left(\frac{S}{\rho}\right) = y + A \cdot K^{(-q)} \,, \tag{6}$$

where the parameters $y$, $A$ and $q$ depend on the material (air or silicon). The following values were used $y_{air} = 1.060$, $A_{air} = 316.5$, $q_{air} = 0.8847$, $y_{Si} = 0.9438$, $A_{Si} = 265.1$, $q_{Si} = 0.8669$.

Equations 5 and 6 allow computing $K_0$ as a function of $K_{avg}$.

The corrections introduced in the previous equations were validated through Monte Carlo simulations using the Geant4 code. The ToF values were obtained by simulating the experimental setup of Figure 7 for the four different distances between the sensors used in the beam test (7, 37, 67, 97 cm), and incident beam energies between 59 to 230 MeV, in intervals of 1 MeV. The incident proton beam energies at the detector entrance were reconstructed from the ToF value determined by using Equation 4, corrected with Equation 5, and benchmarked against the simulated energy. Figure 8 reports the difference between the simulated incident beam energy and the corresponding average reconstructed quantity calculated with and without applying corrections for the energy loss in the telescope, as a function of the beam energy.

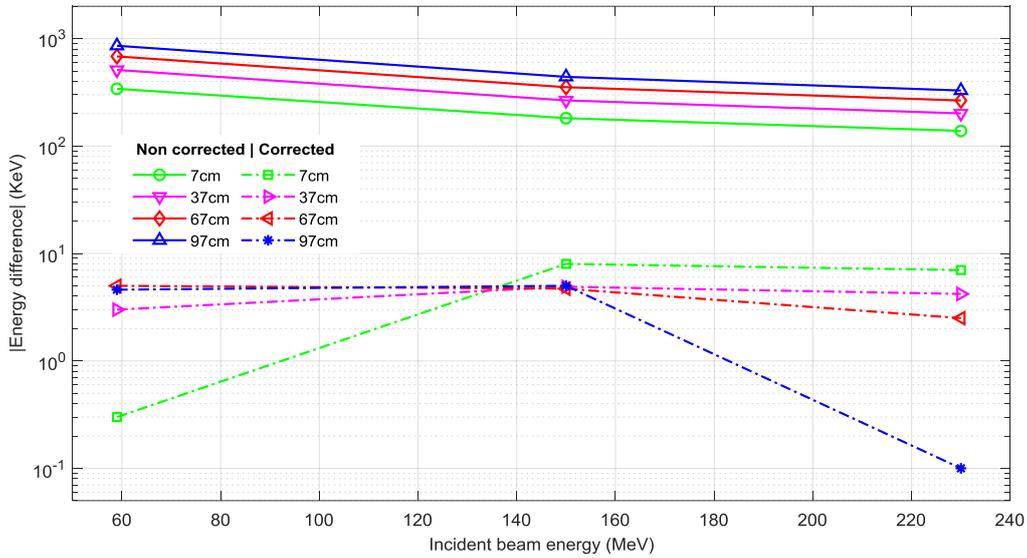

**Figure 8.** Difference between calculated and simulated proton beam energies as a function of the incident beam energy with (dot-dashed lines) and without (continuous lines) correction for the energy loss in air and silicon.

The simulation results confirm that the energy lost by the protons in the material is not negligible especially for the lowest energies and largest distances. The corrections introduced in Equation 5 limit the systematic error at most to a few tens of keV.

## 3. Results

The width of the peak in the distribution of the time difference values, such as the one reported in Figure 6, allows estimating the time resolution for single crossing as $\sigma_{\Delta t}/\sqrt{2}$. Single crossing time resolutions between 75 and 115 ps, depending on the beam energy, were found. These values are well in accordance with the results obtained by other measurements for sensors of 80 μm thickness (25).

The average time difference $\Delta t_{mean}$ measured with the procedure described in Section 2.6 are shown in Figure 9 for the five different beam energies and the four distances between the two sensors of the telescope used in the CNAO test.

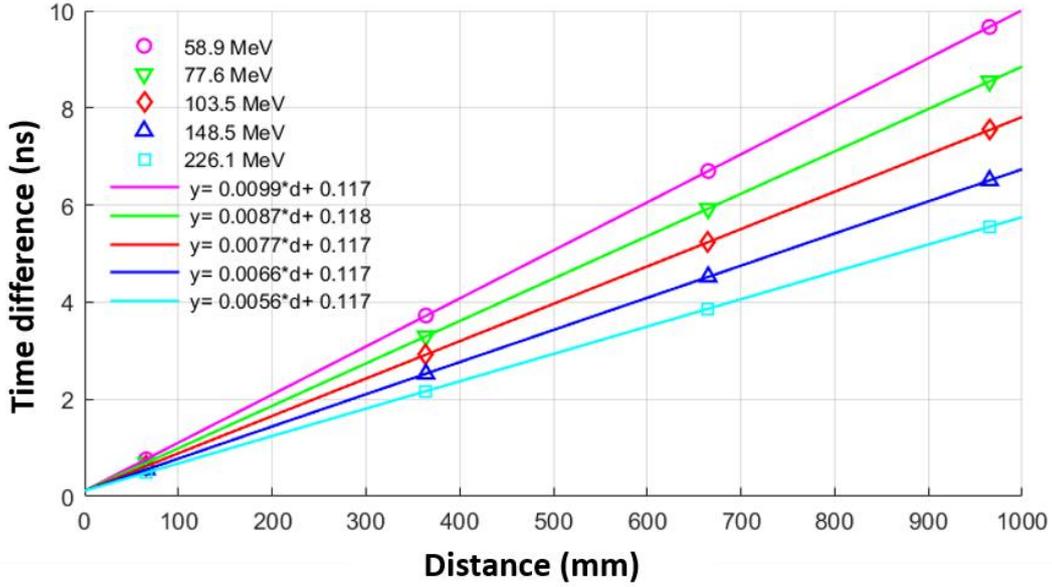

**Figure 9.** Measured average time differences as a function of the distance for 5 proton beam energies (nominal values 58.9, 77.6, 103.5, 148.5, 226.1 MeV) and 4 different distances between the two sensors (7, 37, 67 and 97 cm).

The $\Delta t_{mean}$ values as a function of the distance were linearly interpolated for each beam energy. The slopes of the interpolation lines increase for decreasing energies, being proportional to the inverse velocity. The intercepts of the interpolation lines provide the time offset and came out to be independent of the energy, being essentially induced by differences in the routing of the signals. The statistical error on the measured $\Delta t_{mean}$ is of the order of a few ps, for irradiation times ranging between ~3 s (corresponding to ~5200 coincident signals for d=7 cm between the sensors and 58.9 MeV beam energy) and 9 s (corresponding to ~4700 coincident signals for d=97 cm between the sensors and 226.1 MeV beam energy).

The calibration procedure (Section 2.7) was used to obtain the values of the time offset and of the four relative distances between the sensors, which are reported in Table 1. These parameters and the corresponding uncertainties were then used to calculate the ToF values from the measured $\Delta t_{mean}$ and, using the equations of Section 2.8, the beam energies. The results are reported in Table 2 where the uncertainties on the fit parameters and on $\Delta t_{mean}$ were propagated to the results. The effective acquisition time, corresponding to the fraction of the total time acquired by the digitizer, and the total irradiation time are also reported in Table 2.

**Table 1.** Distances and time offset obtained from the Chi-square minimization method. The latter considered 11 degrees of freedom (dof), resulting from 16 measurements and 5 free parameters (4 distances and 1 time offset). A Chi-square/dof of ~ 5 was obtained.

| Parameter | Value | Error |
|---|---|---|
| d1   (mm) | 65.78 | 0.33 |
| d2   (mm) | 364.02 | 0.42 |
| d3   (mm) | 665.10 | 0.46 |
| d4   (mm) | 965.15 | 0.49 |
| *offset* (ns) | 0.117 | 0.003 |

**Table 2.** List of the measured ToFs and energies with corresponding deviations and statistical errors, along with effective time of acquisition and total acquisition time. The considered distances and time offset (and relative errors) were provided by the Chi-Square minimization (Table 1), while 0.1 MeV was considered as error on the nominal energy values. The values related to the 103.5 MeV nominal energy are highlighted, as they represents the energy value not considered in the calibration method that can be used as an unbiased test point.

| Distance # (mm) | Nominal Energy (MeV) | ToF (ns) | $\sigma_{ToF}$ (ns) | Measured Energy (MeV) | $\sigma_{Emeas}$ (MeV) | Deviation* MeV | $\sigma_{Dev}$ (MeV) | Effective acquisition time (ms) | Total acquisition time (s) |
|---|---|---|---|---|---|---|---|---|---|
| 1 (65.78) | 58.9 | 0.647 | 0.003 | 59.4 | 0.3 | -0.5 | 0.3 | 1.11 | 3.42 |
|  | 77.6 | 0.574 | 0.003 | 77.3 | 0.4 | 0.3 | 0.4 | 1.50 | 4.62 |
|  | *103.5* | *0.507* | *0.003* | *102.8* | *0.6* | *0.7* | *0.7* | *2.41* | *7.42* |
|  | 148.5 | 0.437 | 0.003 | 146.6 | 1.5 | 1.9 | 1.5 | 1.30 | 3.99 |
|  | 226.1 | 0.375 | 0.003 | 218.1 | 3.2 | 8.0 | 3.2 | 1.19 | 3.67 |
| 2 (364.02) | 58.9 | 3.607 | 0.003 | 58.7 | 0.1 | 0.2 | 0.3 | 2.27 | 6.99 |
|  | 77.6 | 3.184 | 0.003 | 77.1 | 0.1 | 0.5 | 0.4 | 2.12 | 6.51 |
|  | *103.5* | *2.804* | *0.003* | *103.0* | *0.2* | *0.5* | *0.4* | *2.77* | *8.53* |
|  | 148.5 | 2.406 | 0.003 | 148.8 | 0.2 | -0.3 | 0.5 | 3.31 | 10.18 |
|  | 226.1 | 2.044 | 0.003 | 228.4 | 0.6 | -2.3 | 0.8 | 2.89 | 8.89 |
| 3 (665.10) | 58.9 | 6.582 | 0.004 | 59.0 | 0.1 | -0.1 | 0.1 | 2.34 | 7.20 |
|  | 77.6 | 5.805 | 0.003 | 77.6 | 0.9 | 0.0 | 0.1 | 2.14 | 6.60 |
|  | *103.5* | *5.114* | *0.003* | *103.6* | *0.1* | *-0.1* | *0.2* | *2.78* | *8.55* |
|  | 148.5 | 4.403 | 0.003 | 148.3 | 0.2 | 0.2 | 0.2 | 2.56 | 7.88 |
|  | 226.1 | 3.746 | 0.003 | 226.4 | 0.4 | -0.3 | 0.4 | 2.93 | 9.01 |
| 4 (965.15) | 58.9 | 9.549 | 0.003 | 59.2 | 0.0 | -0.3 | 0.1 | 2.43 | 7.48 |
|  | 77.6 | 8.432 | 0.004 | 77.6 | 0.1 | 0.0 | 0.1 | 2.21 | 6.81 |
|  | *103.5* | *7.431* | *0.004* | *103.3* | *0.1* | *0.2* | *0.1* | *2.62* | *8.07* |
|  | 148.5 | 6.390 | 0.003 | 148.3 | 0.2 | 0.2 | 0.2 | 2.99 | 9.21 |
|  | 226.1 | 5.440 | 0.003 | 226.0 | 0.3 | 0.1 | 0.304 | 3.01 | 9.25 |

*Difference between the measured energy and the nominal one

It can be observed that, despite the lower system efficiency (see Figure 4), the two larger distances provide a more precise and accurate energy measurement, whereas at the smaller distances the uncertainties and the deviations are larger. According to the detector requirements described in Section 2.3 and, in particular, into Figure 2, the obtained statistical error $\sigma_{ToF}$ is compatible with the clinically acceptable accuracy in the beam energy for the distances 3 and 4, partially compatible (for the lowest energies) for the distance 2 and not compatible for the distance 1. More specifically, for the run at 97 cm distance and maximum beam energy, 6 s of irradiation (out of the 9 acquired in total) allow measuring ~3000 coincident signals, which are sufficient to reach the required statistical error on the ToF.

Figure 10 shows the deviations between calculated and nominal energies for the tests at 67 and 97 cm distances and for the five considered energies. They are found to be always smaller than 0.5 MeV. In particular, the error of the test point (i.e. 103.5 MeV) is ~150 keV for both considered distances.

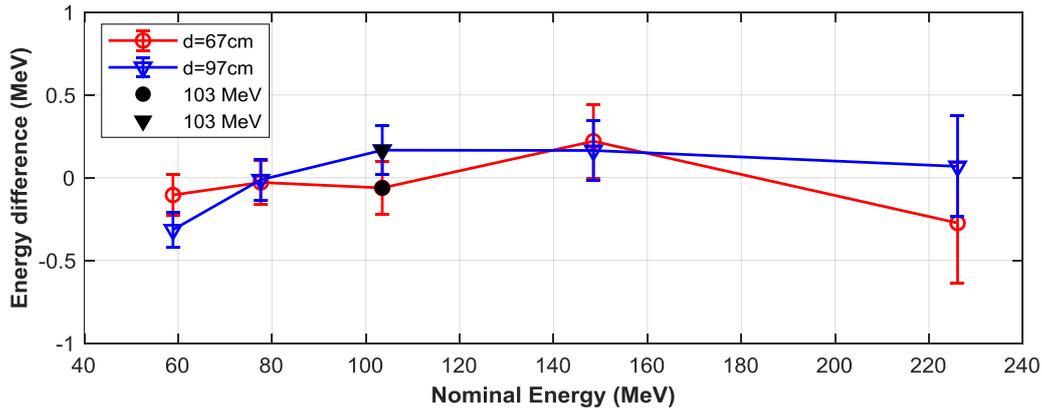

**Figure 10.** Deviations between the measured and the nominal energy for the five different energies (58.9, 77.6, 103.5, 148.5, 226.1 MeV) and two distances (67 and 97 cm). The test point at 103.5 MeV is marked in black.

Although the energy deviation is an interesting parameter for the evaluation of the detector and method accuracies, the corresponding range deviation in water is more important from the clinical point of view. Therefore, all proton energies were converted into water ranges using a four-degree polynomial formula generated using the empirical Bragg-Kleman rule (33,34) and ICRU-49 data (35) as reported by (36). The differences in water range between nominal and measured proton energies at 67 cm and 97 cm are shown in Figure 11. The range discrepancies remained within half millimetre for the lower energies and within one millimetre for the maximum energy, complying with the clinical requirements.

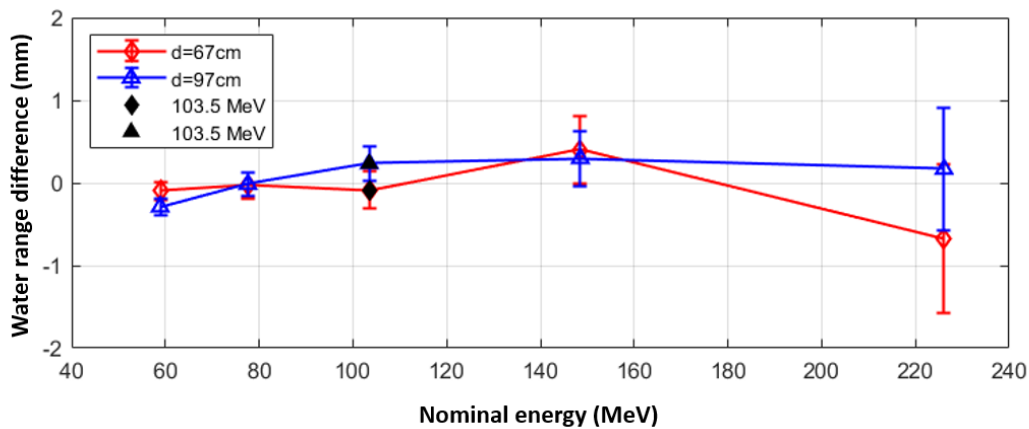

**Figure 11.** Water range differences as a function of the proton beam energy. The test point at 103.5 MeV is marked in black

## 3 Discussion

The online energy check of proton pencil beams is gaining increasing interest among the particle therapy community, because it would support and enhance the development of new and robust dose delivery techniques able to fast change the beam energy during the treatment. Indeed, 4D approaches for the treatment of moving targets, such as tumor tracking, would greatly benefit of such improvements.

This work describes a prototype detector made by a telescope of two UFSD pads and the results of its test at the CNAO clinical proton beam, with an energy range between 59 and 226 MeV. The goal of this contribution is to demontrate the proof-of-concept of a new detector able to directly and accurately measure the proton beam energy during the treatment.

The method used to derive the beam energy from the measured ToF of protons in a telescope of two UFSD sensors was considering the particles energy loss in the sensors and in air, and Monte Carlo simulations were perfomed to verify the adopted approach.

The energy measurement aims at assessing the beam range, which is the paramount parameter in clinics. The clinically acceptable range uncertainty (< 1 mm) determines the needed detector performances in terms of accuracy and time resolution. As an example, the maximum error on the ToF measurement for a proton beam energy of 230 MeV with 1 m distance between the sensors must not exceed 4 ps for an error below 1 mm in the range in water.

During the test at CNAO, six seconds of irradiation with protons at $5 \times 10^8$ p·s$^{-1}$ intensity were necessary to collect the statistics required to keep the maximum acceptable statistical error below the aforementioned tolerance. However, the system development should proceed towards the future goal of the energy measurement during treatment, ideally within the spot duration (ranging from tens of ms to hundreds of ms), and therefore a reduction of the irradiation time needed to obtained the required accuracy is mandatory. It is worth noticing that the effective acquisition time used in the test is of 2-3 ms and that digitizer dead time (~600 µs) was the main bottleneck to set the irradiation time needed to collect the proper statistics. The authors are aware that an optimization of the data acquisition and throughput could lead to a reduction of total acquisition time. On the other side, sensors dedicated to this project have been designed and produced and will be tested on clinical beams in the next future. These sensors, segmented in strips, will allow the simultaneous acquisition of several channels (thanks to a dedicated redout electronics under development), enlarging the sensisitive area and, consequently, the statistics collection of coincident proton signals.

The UFSD pads of the telescope detector feature an active area of 80 µm thickness and showed a time resolution of ~75-115 ps in the tests performed at CNAO. Since the time resolution of UFSD sensors can be improved up to few tens of ps by reducing the sensor thickness (25), the new sensors feature a 50 µm thickness and will probably reduce by a factor 4 the needed irradiation time. Moreover, the total thickness of the dedicated sensors will be thinned to ~100 µm to further minimize the beam perturbation and increase the system efficiency.

The results demonstrated that the statistical errors at 67 and 97 cm distance between the sensors of the telescope are compatible with the accuracy requirements on the ToF measurement, and allow achieving the range accuracy clinically recommended.

Although promising, this study has some limitations. The systematic errors depending on the experimental setup, i.e. distance and time offset, have been studied, but need to be further investigated, together with other possible source of systematic errors, when a stiff and light sophisticated mechanical support will be developed to allow the precise positioning of the sensors positions. This future mechanical support will also introduce the possibility of rotating the detector in order to irradiate the telescope from both sides, thus directly measuring the time *offset* contribution.

Moreover, the global fit approach adopted for the detector calibration was necessary because of the limited knowledge of the detector distances. It relies on several measurements and on the knowledge a priori of few beam energies used as calibration points. A new approach will be considered during the second part of the project to provide an absolute energy measurement, i.e. independent of the nominal energy values provided by the clinical facility. Last, although it is worth noticing that no degradation of sensor performances were recorded during the tests, studies dedicated to radiation hardness were not yet performed and will be investigated in the second part of the project. These will surely benefit of the increasing knowledge of the research community about the radiation resistance of UFSD sensors (37).

# 4 Conclusion

A prototype detector able to measure in a few seconds the clinical proton beam energy with the ToF technique was built and tested in the CNAO clinical center. The UFSD technology was used to reach the time resolution necessary to measure the range in water with an error < 1 mm. The energy measurements have been benchmarked with both Monte Carlo simulations and the data from the commissioning of the CNAO clinical beam. Although considering the needed improvements in terms of mechanical support, detector sensitive area, readout electronics, data acquisition and throughput, the reported results demonstrate the feasibility of a fast and online energy measurement tool of clinical proton beams.

**Acknowledgments**

The authors have no relevant conflict of interest to disclose.

This work has been performed within the framework of the MoVeIT project, founded by the INFN-CSN5, and has been supported by MIUR Dipartimenti di Eccellenza (ex L.232/2016, art.1, cc. 314, 337). FMM was supported by the FAPESB fellowship. Part of this work has been also supported by the European Union's Horizon 2020 Research and Innovation funding program (Grant Agreement no. 669529 - ERC UFSD669529) and by Hamamatsu Photonics K.K. (HPK).

**REFERENCES**

1. Chang JY, Zhang X, Knopf A, Li H, Mori S, Dong L, et al. Consensus Guidelines for Implementing Pencil-Beam Scanning Proton Therapy for Thoracic Malignancies on Behalf of the PTCOG Thoracic and Lymphoma Subcommittee. Int J Radiat Oncol. 2017 Sep;99(1):41–50.

2. Lomax A. Intensity modulated proton therapy and its sensitivity to treatment uncertainties 1: the potential effects of calculational uncertainties. Phys Med Biol. 2008 Feb;53(4):1027–42.

3. Giordanengo S, Manganaro L, Vignati A. Review of technologies and procedures of clinical dosimetry for scanned ion beam radiotherapy. Phys Medica. 2017 Nov;43:79–99.

4. Ferrero V, Fiorina E, Morrocchi M, Pennazio F, Baroni G, Battistoni G, et al. Online proton therapy monitoring: clinical test of a Silicon-photodetector-based in-beam PET. Sci Rep. 2018 Dec;8(1):4100.

5. Golnik C, Hueso-González F, Müller A, Dendooven P, Enghardt W, Fiedler F, et al. Range assessment in particle therapy based on prompt $\gamma$-ray timing measurements. Phys Med Biol. 2014;59(18):5399–422.

6. Knopf A-C, Lomax A. In vivo proton range verification: a review. Phys Med Biol. 2013;58(15):R131--R160.

7. Graeff C, Lüchtenborg R, Eley JG, Durante M, Bert C. A 4D-optimization concept for scanned ion beam therapy. Radiother Oncol. 2013;109(3):419–24.

8. Mazzucconi D, Agosteo S, Ferrarini M, Fontana L, Lante V, Pullia M, et al. Mixed particle beam for simultaneous treatment and online range verification in carbon ion therapy: Proof-of-concept study. Med Phys. 2018;45(11):5234–43.

9. Grözinger SO, Bert C, Haberer T, Kraft G, Rietzel E. Motion compensation with a scanned ion beam: a technical feasibility study. Radiat Oncol. 2008;3(1):34.

10. Psoroulas S, Bula C, Actis O, Weber DC, Meer D. A predictive algorithm for spot position corrections after fast energy switching in proton pencil beam scanning. Med Phys. 2018 Nov;45(11):4806–15.

11. Giordanengo S, Palmans H. Dose detectors, sensors, and their applications. Med Phys. 2018 Nov;45(11):e1051–72.

12. Mirandola A, Molinelli S, Vilches Freixas G, Mairani A, Gallio E, Panizza D, et al. Dosimetric commissioning and quality assurance of scanned ion beams at the Italian National Center for Oncological Hadrontherapy. Med Phys. 2015;42(9):5287–300.

13. Gillin MT, Sahoo N, Bues M, Ciangaru G, Sawakuchi G, Poenisch F, et al. Commissioning of the discrete spot scanning proton beam delivery system at the University of Texas M.D. Anderson Cancer Center, Proton Therapy Center, Houston. Med Phys. 2010;37(1):154–63.

14. Rana S, Bennouna J, Samuel EJJ, Gutierrez AN. Development and long-term stability of a comprehensive daily QA program for a modern pencil beam scanning (PBS) proton therapy delivery system. J Appl Clin Med Phys. 2019 Apr 1;20(4):29–44.

15. Fassi A, Seregni M, Riboldi M, Cerveri P, Sarrut D, Ivaldi GB, et al. Surrogate-driven deformable motion model for organ motion tracking in particle radiation therapy. Phys Med Biol. 2015;60(4):1565–82.

16. Traini G, Battistoni G, Bollella A, Collamati F, De Lucia E, Faccini R, et al. Design of a new tracking device for on-line beam range monitor in carbon therapy. Phys Medica Eur J Med Phys. 2017 Feb;34:18–27.

17. Graeff C. Motion mitigation in scanned ion beam therapy through 4D-optimization. Phys Medica. 2014;30(5):570–7.

18. Bert C, Saito N, Schmidt A, Chaudhri N, Schardt D, Rietzel E. Target motion tracking with a scanned particle beam. Med Phys. 2007 Nov;34(12):4768–71.

19. van de Water S, Kreuger R, Zenklusen S, Hug E, Lomax a J. Tumour tracking with scanned proton beams: assessing the accuracy and practicalities. Phys Med Biol. 2009;54(21):6549–63.

20. Riboldi M, Orecchia R, Baroni G. Real-time tumour tracking in particle therapy: Technological developments and future perspectives. Vol. 13, The Lancet Oncology. 2012. p. e383--91.

21. Rabin N V. Status and possibilities of the time-of-flight measurement technique using long scintillation counters with a small cross section (Review). Instruments Exp Tech. 2007 Sep;50(5):579–638.

22. H.-W. Sadrozinski et al. Ultra-fast silicon detectors. Nucl Instruments Methods Phys Res Sect A Accel Spectrometers, Detect Assoc Equip. 2013 Dec;730:226–31.

23. Sola V, Arcidiacono R, Bellora A, Cartiglia N, Cenna F, Cirio R, et al. Ultra-Fast Silicon Detectors for 4D tracking. J Instrum. 2017 Feb;12(02):C02072–C02072.

24. Pellegrini G, Fernández-Martínez P, Baselga M, Fleta C, Flores D, Greco V, et al. Technology developments and first measurements of Low Gain Avalanche Detectors (LGAD) for high energy physics applications. Nucl Instruments Methods Phys Res Sect A Accel Spectrometers, Detect Assoc Equip. 2014 Nov 21;765:12–6.

25. Sola V, Arcidiacono R, Boscardin M, Cartiglia N, Betta GD, Ficorella F. Nuclear Inst . and Methods in Physics Research , A First FBK production of 50 μm ultra-fast silicon detectors. Nucl Inst Methods Phys Res A. 2019;924(June 2018):360–8.

26. DT1470ET - CAEN - Tools for Discovery [Internet]. [cited 2020 Mar 6]. Available from: https://www.caen.it/products/dt1470et/

27. CIVIDEC Instrumentation - CVD Diamond Technology applications [Internet]. [cited 2020 Mar 6]. Available from: https://cividec.at/electronics-C2.html

28. DT5742 - CAEN - Tools for Discovery [Internet]. [cited 2020 Mar 6]. Available from: https://www.caen.it/products/dt5742/

29. Agostinelli S et. al. Geant4—a simulation toolkit. Nucl Instruments Methods Phys Res Sect A Accel Spectrometers, Detect Assoc Equip. 2003;506:250–303.

30. Molinelli S, Mairani A, Mirandola A, Vilches Freixas G, Tessonnier T, Giordanengo S, et al. Dosimetric accuracy assessment of a treatment plan verification system for scanned proton beam radiotherapy: One-year experimental results and Monte Carlo analysis of the involved uncertainties. Phys Med Biol. 2013;58(11):3837–47.

31. Vignati A, Monaco V, Attili A, Cartiglia N, Donetti M, Mazinani MF, et al. Innovative thin silicon detectors for monitoring of therapeutic proton beams: preliminary beam tests. J Instrum. 2017 Dec;12(12):C12056–C12056.

32. Tawfeek HM, Hady FM, Jassim MK. Investigations into the Proton Stopping Power of Human Body. J Chem Biol Phys Sci. 2015;5(4):4345–54.

33. Bortfeld T. An analytical approximation of the Bragg curve for therapeutic proton beams. Med Phys. 1997;24(12):2024–33.

34. Ishkhanov BS. The atomic nucleus. Moscow Univ Phys Bull. 2012;67(1):1–24.

35. ICRU 49. Stopping Power and Ranges for Protons and Alpha Particles. Int Comm Radiat Units Meas Rep No 49. 1993;

36. Krim M, Harakat N, Khouaja A, Inchaouh J, Mesradi MR, Chakir H, et al. Method for range calculation based on empirical models of proton in liquid water: Validation study using Monte-Carlo method and ICRU data. Int J Sci Eng Res. 2017;8(3):728–35.

37. Ferrero M, Arcidiacono R, Barozzi M, Boscardin M, Cartiglia N, Betta GFD, et al. Radiation resistant LGAD design. Nucl Instruments Methods Phys Res Sect A Accel Spectrometers, Detect Assoc Equip. 2019 Mar 1;919:16–26.